\documentclass[aps,prl,twocolumn,showpacs,groupedaddress]{revtex4}
\usepackage{txfonts}  
\usepackage{graphicx}  
\usepackage{bm}        
\usepackage{amsfonts}
\usepackage{amssymb}

\begin{document}



\title{Analytic prediction for planar turbulent boundary layers}
\author{Xi Chen}
\author{Zhen-Su She}
\thanks{she@pku.edu.cn}
\affiliation{State Key Laboratory for Turbulence and Complex Systems
and Department of Mechanics, College of
Engineering, Peking University, Beijing 100871, China}

\date{\today}

\begin{abstract}
Analytic predictions of mean velocity profile (MVP) and streamwise ($x$) development of related integral quantities
are presented for flows in channel and turbulent boundary layer (TBL), based on a symmetry analysis of eddy length
and total stress. Specific predictions are the friction velocity $u_\tau$:
${ U_e/u_\tau }\approx 2.22\ln Re_x+2.86-3.83\ln(\ln Re_x)$; the boundary layer thickness $\delta_e$:
$x/\delta_e \approx 7.27\ln Re_x-5.18-12.52\ln(\ln Re_x)$; the momentum thickness Reynolds number: $Re_x/Re_\theta=4.94[{(\ln {{\mathop{\rm Re}\nolimits} _{\theta} } + 1.88)^2} + 1]$, all in good agreement with empirical data.

\end{abstract}

\pacs{47.27.-i, 47.27.N-, 47.27.E-} \maketitle

Planar turbulent boundary layer (channel and TBL) is a canonical wall-bounded flow of significant theoretical
and practical interests \cite{Voyage}. It is widely seen in the atmospheric flows near the ground,
or the flows over the surface of a vessel, or the wings of an aircraft, \emph{etc}. \cite{SmitsMarusic2013}. Common to all these flows is a thin-layer
close to the wall - named turbulent boundary layer \cite{Prandtl} - where most of flow momentums and energies are dissipated by viscous drag and turbulent fluctuations. The latter two effects have been studied since Reynolds \cite{Reynolds}, using
the ensemble averaged Navier-Stokes equation:
\begin{equation}\label{eq:SWT}
\nu \partial_y U - \left\langle {u'v'} \right\rangle =\tau
\end{equation}
where $U$ is the streamwise mean velocity, $u'$ and $v'$ are streamwise and
normal fluctuating velocity, respectively, $\nu$ the molecular viscosity and $\tau$ the total stress representing
the driving force (defined later).
This equation is unclosed in the presence of fluctuating velocities, and the long-standing question is how to
predict $U$ for all Reynolds numbers, and then the streamwise development of other relevant quantities
(including the friction coefficient) \cite{Voyage}. Until today, it is still a vivid challenge to theorists \cite{Marusic2010}, despite over a century's effort since the seminar work of Prandtl in 1904 proposing the concept of the boundary layer \cite{Prandtl}.

In modern analysis, the difficulty of the problem is particularly attributed to the interplay of two different characteristic scales \cite{Yakhot2010}. Denote $S = \partial_y U$, the mean shear,
$W = - \left\langle {u'v'} \right\rangle $ the turbulent Reynolds shear stress.
When the leading balance of (\ref{eq:SWT}) is between $\nu S$ and $\tau$, which
occurs in the inner flow region close to the wall, the friction velocity ($u_\tau= \sqrt{\tau_w}$, $\tau_w$ the wall stress) and
viscous length ($\ell_\nu=\nu/u_\tau$) define the correct scales exhibiting data
collapse of $U$ at different $Re$'s, known as the {law of wall}. However, for outer flow ($\ell_\nu\ll y\le \delta_e$, $\delta_e$ the boundary layer thickness commonly set as $\delta_{99}$)
where $\nu S\ll W$, choosing $u_\tau$ or $U_e$ (freestream velocity) as the velocity scale would respectively lead
to log law \cite{millikan1938} or power law \cite{Barenblatt,George06} description of $U$ in an intermediate overlap region for asymptotically large $Re$'s.
The debate between the two proposals continues to attract attention \cite{Marusic2010}, for both allow a satisfactory description of data in
their restricted flow regions. On the other hands, more empirical approaches \cite{Nickels04,monkewitz2007}
introduce composite formulas describing the mean velocity for the entire domain, but with many free parameters
of purely fitting. In short, analytic approach with clear physical picture and testable assumptions is
missing, leaving important questions as the universality of Karman constant, and effects of geometry (internal
versus external flows) unaddressed.

Here, we propose a symmetry-based approach yielding a closure solution for the mean momentum and kinetic energy
equations. Our analysis begins with an symmetry analysis of a characteristic length of
energy containing eddy, which enables a prediction of its functional form. It also quantifies the geometry difference between channel and TBL, through different approximations of kinetic energy
equation as well as the total stress, thus going beyond a recent theory of channel by L'vov \emph{et al.} \cite{Lvov2008} (here refer to as LPR). The predicted mean velocity
covers the entire outer flow domain; and in the case of TBL, the streamwise
development of a series of global quantities, i.e. friction coefficient, shape factor, boundary layer and momentum thicknesses, are predicted, agreeing well with data. Compared to previous Pades approximations \cite{monkewitz2007} (here refer to as MCN) and multilayer models
\cite{Nickels04}, the current theory involves only three parameters for TBL (and for channel), which are
$Re$-independent; in particular, the Karman constant $\kappa\approx 0.45$ is remarkably universal in channel and TBL
(also applied to pipes \cite{shenjp}). The result resolves the debate on the scaling of mean velocity, in favor of the log law,
with corrections beyond the leading order in $1/\ln Re_\tau$.

For the flow over a flat plate ($0\leq x \leq \infty$ and $y=0$), the Navier-Stokes-Prandtl equations read:
\begin{eqnarray}\label{MME0}
{{\partial_x U}}+ {{\partial_y V}}&=& 0 \\
 U{{\partial_x U}}+ V{{\partial_y U}}+\partial_x P&=& \nu {{{\partial_y \partial_y}U}} - {\partial_y }{{}}\left\langle {u'v'} \right\rangle \label{MME1}
\end{eqnarray}
where $V$ the mean vertical velocity (zero in channel) and $\partial_x P$ the mean pressure gradient (zero in TBL).
Integrating (\ref{MME1}) in $y$ yields (\ref{eq:SWT}), where for channel $\tau=u_\tau^2 r'$ ($r'=r/\delta_e$ and $r=\delta_e-y$; $\delta_e$ the half height of a channel), and for TBL $\tau= u_\tau^2+\int_0^y (U\partial_x U+V\partial_{y'} U)dy'$. Substituting
(\ref{MME0}) into (\ref{MME1}) and integrating the latter from 0 to $\delta_e$, one obtains the von Karman's integral
momentum equation for TBL:
\begin{equation}\label{eq:Cf}
{{d\theta }}/{{dx}} = u_\tau^2/{{U_e^{2}}}=C_f/2 
\end{equation}
where $\theta  = \int_0^{{\delta _e}} {U/{U_e}(1 - U/{U_e})} dy$ is the momentum thickness, and $C_f=2u_\tau^2/{{U_e^{2}}}$
is the friction coefficient (here we set zero $S$ and $W$ for $y\geq\delta_e$). Moreover, the turbulent kinetic energy is
described by the mean kinetic energy equation, i.e.
\begin{equation}\label{eq:MKED}
        {S}{W} + {\Pi } = {\varepsilon},
\end{equation}
where $\mathcal{P}=SW$ is the production; $\Pi$ represents the spatial energy transfer (including diffusion, convection and
fluctuation transport); $\varepsilon$ is the viscous dissipation.
Note that (\ref{eq:SWT}) and (\ref{eq:MKED}) describe the two fundamental processes in the flow, i.e. momentum and energy
transports, respectively; and (\ref{eq:Cf}) characterizes the streamwise scaling of friction coefficient
for TBL. We now construct a closure solution as below.

Note that according to (\ref{eq:MKED}), $\varepsilon$ should be determined by $S$ and $W$, and a characteristic length $\ell$ describing the effects
of spatial energy transfer $\Pi$. Thus, a dimensional argument yields $\varepsilon (S,W,\ell) = W^{(1+\frac{n}{2})}S^{(1-n)}\ell^{(-n)}$ or
\begin{equation}\label{eq:elln}
\ell= W^{(\frac{1}{n}+\frac{1}{2})} S^{(\frac{1}{n}-1)} \varepsilon^{(-\frac{1}{n})},
\end{equation}
where $n$ is an arbitrary real number. Note that as $n \rightarrow \infty$, the
dissipation drops out, and the resulting length becomes the classical
mixing length of Prandtl \cite{Prandtl}: $\ell_\infty=\sqrt{W}/S$. Furthermore, for a channel flow, ${S }\propto r$ due to the central mirror symmetry, and (\ref{eq:SWT}) can be approximated
to ${W} \approx \tau\propto r $. Meanwhile, $\varepsilon$ tends to $\varepsilon_0>0$ as $r\rightarrow0$.
Thus, (\ref{eq:elln}) yields $\ell\propto r^{\frac{2}{n}-\frac{1}{2}}$, where only $n=4$ results in a finite nonzero
value $\ell_0$ at $r=0$. Hence a unique characteristic length $\ell_\varepsilon$ is defined as
\begin{equation}\label{eq:elleps}
 \ell_\varepsilon= {W^{\frac{3}{4}}}{S^{\frac{-3}{4}}}\varepsilon^{\frac{-1}{4}}.
\end{equation}
This length recovers the crucial scaling function in the LPR model \cite{Lvov2008} without involving any wall function.

For TBL, although there is no central symmetry by the opposite wall, $\ell_\varepsilon$ can still be defined
in the same way as in channel by setting $n=4$ in (\ref{eq:elln}). A notable fact is that, the ratio between
$\ell_\varepsilon$ and any other length of different $n$ in (\ref{eq:elln})
is exactly related to the dissipation production ratio, $\Theta\equiv{\varepsilon}/ {(SW)}$, namely
${\ell_{n}}/{\ell_{\varepsilon}}=(\frac{\varepsilon}{SW})^{\frac{1}{4}-\frac{1}{n}}= \Theta^{\frac{1}{4}-\frac{1}{n}}$,
Interestingly, when $\Theta\approx1$ corresponding to the well-known quasi-balance
regime (QBR), all lengths reduce to a single one, namely $\ell_\varepsilon$. For TBL, QBR extends from the wall
(above the buffer layer) to the boundary layer edge. In contrast, for channel, near the centerline (where we call the core layer),
$\Theta\propto r^{-2}$, and all lengths are different. A simple matching argument can derive an expression for $\Theta$,
valid from QBR to central core. Assume $\Theta\approx c/r'^{2}$ as $r'\to 0$,
where $c$ is a dimensionless coefficient. To match $\Theta\to 1$ as $r'\to 1$,
a composite expression is $\Theta  = 1 +  c/r'^{2}- {c}$, which can be rewritten as
\begin{equation}\label{eq:theta}
\Theta(r')=[{1 + {({r_{c}}/r')^2}}]/{(1 + r_{c}^2)}
\end{equation}
where ${r_{c}} = \sqrt {{c}/(1 - {c})} $ indicates the thickness of the core layer (given later).
A variant of (\ref{eq:theta}) is for the spatial energy transfer in (\ref{eq:MKED}), that is
$\Pi = \mathcal{P}(\Theta  - 1)\propto(1 - r'^2)$. The parabolic form stems from the simplest expansion
at $r=0$ in the presence of the central mirror symmetry,
namely, $\partial_r \Pi=0$. Hence, (\ref{eq:theta}) is a reasonable approximation, which
solves (\ref{eq:MKED}). Note that (\ref{eq:theta}) also applies to TBL, where $r'=\delta_e-y$ the distance to the
boundary layer edge, and $r_c=0$ (hence $\Theta\equiv 1$).

\begin{figure*}
\includegraphics[trim = 3cm 1.5cm 3.7cm 2.2cm, clip, width = 5.8 cm]{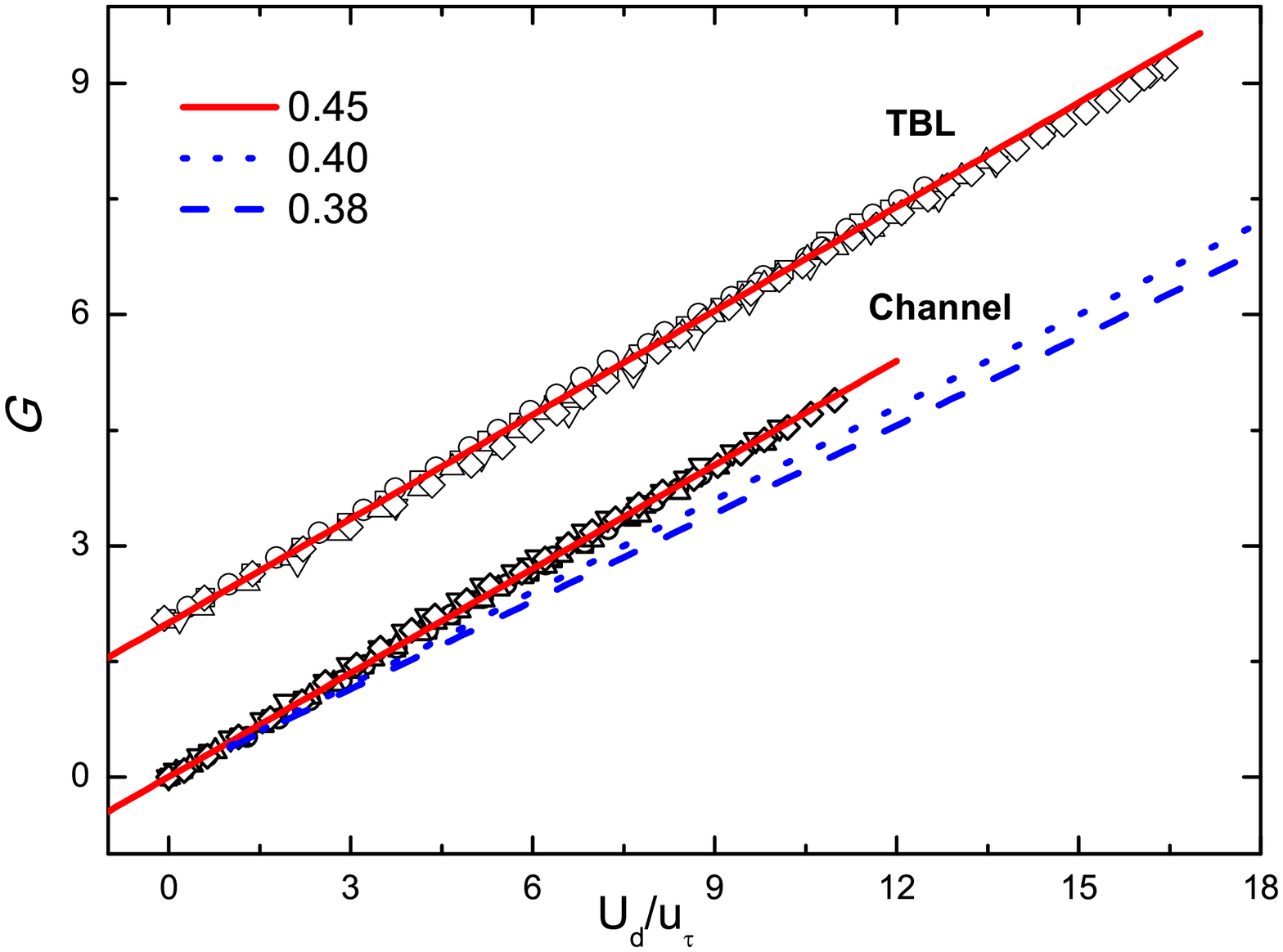}   
\includegraphics[trim = 3cm 1.5cm 3.7cm 2.2cm, clip, width = 5.8 cm]{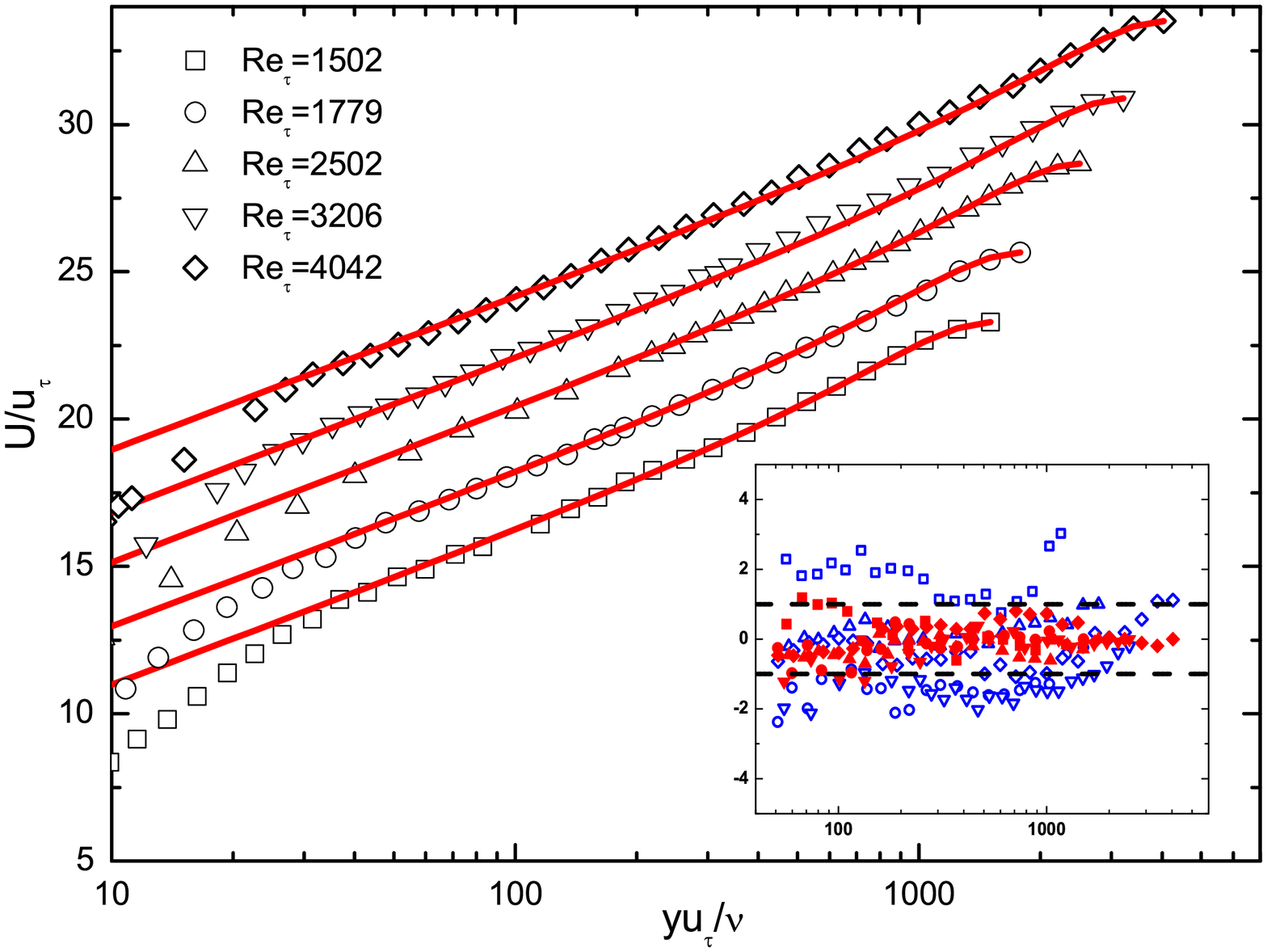}
\includegraphics[trim = 3cm 1.4cm 3.7cm 2.2cm, clip, width = 5.8 cm]{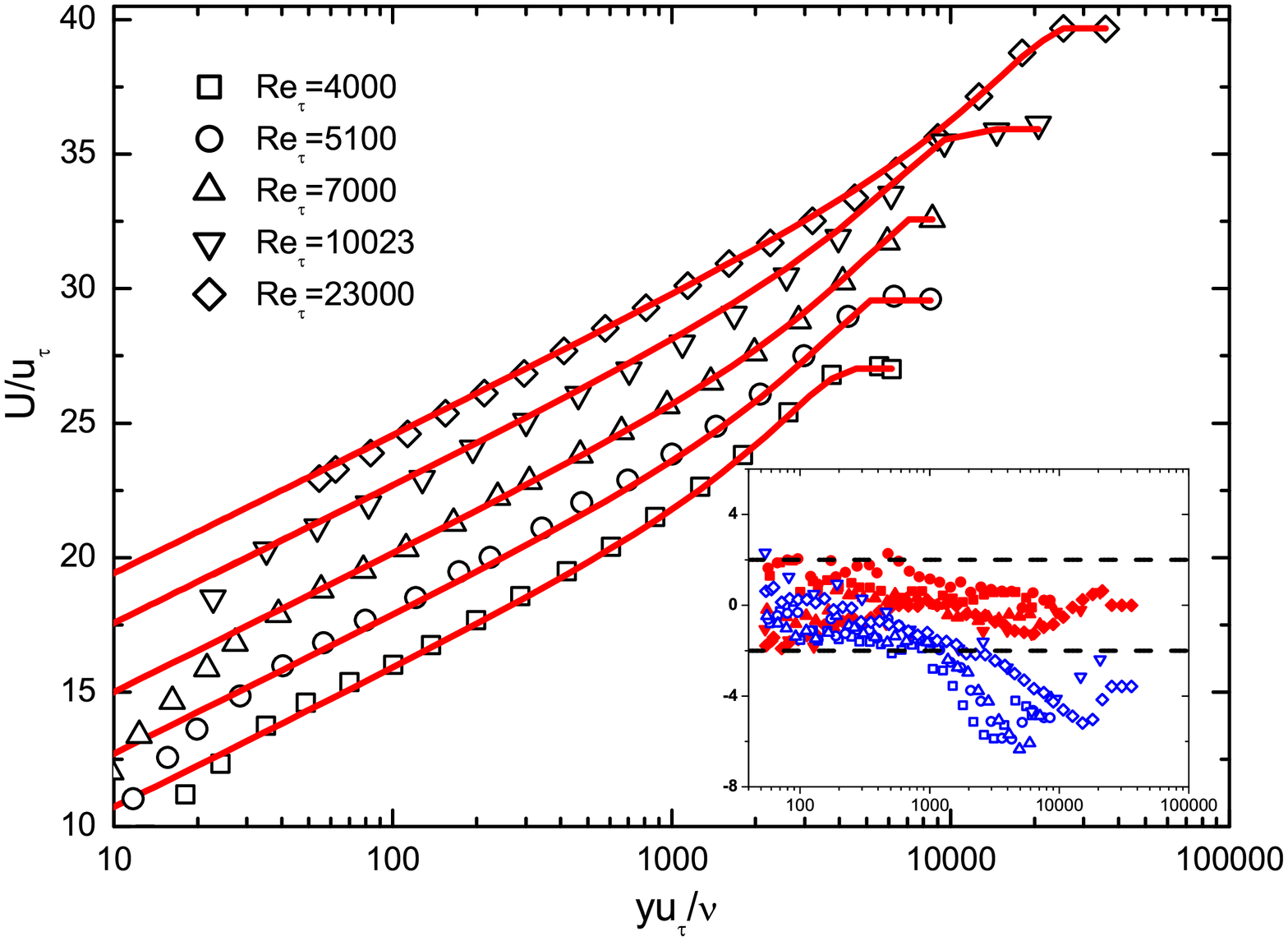}
    \caption{(color online). [Left] Validation of (\ref{eq:Ud2}) by experimental $U_d/u_\tau$ at large $Re$'s. Solid lines indicate the universal slope $\kappa=0.45$, dotted line $\kappa=0.40$ and dashed line $\kappa=0.38$. [Middle] Predictions (lines) of MVP for channel. Inset shows the relative errors $100\times(U^{EXP}/U^{Theory}-1)$,
    where red solids are current predictions bounded within 1\% (dashed lines), and blue opens are the LPR model \cite{Lvov2008}. [Right] Predictions (lines) of MVP for TBL. Red solids in the inset are current predictions within 2\% (dashed lines), and blue opens are
the MCN model \cite{monkewitz2007}. Profiles are vertically staggered for clarity. Channel: all data from \cite{Monty2005};
    TBL: $Re_\tau=4000,5100,7000$
    from \citep{Carlier}; $10023$ from \citep{Degraaff2000}; 23000 from \citep{Nickels2007}).}
    \label{fig:Ell}
\end{figure*}

An interpretation of $\ell_\varepsilon$ is inspired by the fact that $\ell_\varepsilon$ can
also be expressed in terms of the eddy viscosity $\nu_t=W/S$, i.e. $\ell_\varepsilon=\nu_t^{3/4}/\varepsilon^{1/4}$,
which can be contrasted to the Kolmogorov dissipation length $\eta=\nu^{3/4}/\varepsilon^{1/4}$. Since $\eta$ represents the scale
of the smallest eddies dissipating energy, we interpret $\ell_\varepsilon$ to correspond to the
energy containing eddies initiating the energy transfer from the mean shear to turbulence.

Now, we develop a symmetry argument to derive the functional form of $\ell_\varepsilon$ and $\tau$. It consists in postulating a scaling form originated from a dilation invariance in the direction
normal to the wall. The key element is the dilation center, or the fixed point for dilation transformation group,
which is located at $r=0$ for $\ell_\varepsilon$ where $\ell_\varepsilon=\ell_\varepsilon^{max}=\ell_0$,
and at $y=0$ for $\tau$ where $\tau=\tau^{max}=\tau_w$. In both cases, we assume that the dilation invariant form is $d\ell_\varepsilon/dr\propto r^\alpha$ and $d\tau/dy\propto y^\beta$, since $d\ell_\varepsilon/dr =0$ at $r'=0$ and $d\tau/dy =0$ at $y=0$.

For $\ell_\varepsilon$, the boundary condition at the wall, i.e. $\ell_\varepsilon=0$
at $r'=1$, yields a defect power law form: $\ell_\varepsilon/\ell_0= 1-r'^m$, with $m=\alpha+1$
a free parameter characterizing the scaling for the bulk flow.
Presently, we set $m=4$, whose rationality will be discussed elsewhere. Note that close to the wall ($r'\rightarrow1$),
$1-r'^4\approx 4y/\delta_e$ and $\ell_\varepsilon\approx 4y(\ell_0/\delta_e)$. The latter is consistent with the recent
studies that the scale of large eddies is proportional to the wall distance \cite{Jimenez2012}.
Denoting $4\ell_0/\delta_e=\kappa$ (so that $\ell_\varepsilon\approx \kappa y$ when $r'\rightarrow1$),
the form of $\ell_\varepsilon$ is finally:
\begin{equation}\label{eq:elleps2}
 \ell_\varepsilon/\delta_e= \kappa(1-r'^4)/4 .
 \end{equation}
We emphasize that (\ref{eq:elleps2}) applies to both channel and TBL under the same planar wall condition.

Similarly, for $\tau$, we obtain
\begin{equation}\label{eq:tau}
  {\tau }/{\tau_w} = 1-(y/\delta_e)^\gamma = 1-{(1-r')^\gamma },
\end{equation}
where $\gamma=1+\beta$ is also an exponent to be determined. For channel, we have an exact result:
$\tau={\tau_w} r'$, hence (\ref{eq:tau}) is rigorous with $\gamma=1$. For TBL, it has been noticed that $\gamma \ne 1$ \cite{Degraaff2000}. Here, we argue that $\gamma>1$, as it would signify a larger magnitude of the Reynolds stress
(since $W\approx \tau$),
hence also a larger turbulent production ($SW$) as observed by Jimenez \emph{et al.} \cite{Jimenez2010}.
Inspecting DNS data \cite{schlatter2010}, we propose an empirical $\gamma=3/2$ for TBL, which keeps invariant for all $Re$'s.

Substituting $\Theta=\varepsilon/(SW)$ into (\ref{eq:elleps}) yields $S=\sqrt{W}/(\ell_\varepsilon\Theta^{1/4})$.
Integrating $S$ in $r$ and using $W\approx{\tau}$, one has
\begin{equation}\label{eq:Ud}
U_e-U=\int_0^r Sd\hat{r}\approx \int_0^r \frac{\sqrt{\tau}}{\ell_\varepsilon\Theta^{1/4}} d\hat{r}.
\end{equation}
Substituting (\ref{eq:theta}), (\ref{eq:elleps2}) and (\ref{eq:tau}) into (\ref{eq:Ud}) yields:
\begin{eqnarray}\label{eq:Ud2}
(U_e-U)/u_\tau\approx \frac{1}{\kappa}\int_0^{r'} {f(\hat{r})} d\hat{r},
\end{eqnarray}
where \[f(\hat{r})=\frac{{4 \sqrt{{1 - {{(1 - \hat{r})}^{\gamma}}}} {{(1 + {r_c}^2)}^{1/4}} }}
{[1+({r_c/\hat{r}})^2]^{1/4}(1 - \hat{r}{^4})}. \]
Denoting $G(r')=\int_0^{r'} {f(\hat{r})} d\hat{r}$, (\ref{eq:Ud2}) yields $G=\kappa U_d/u_\tau$ where $U_d=U_e-U$.
This linear relation is indeed validated with high accuracy (Fig.\ref{fig:Ell} left), where $r_c=0$ and
$\gamma=3/2$ (empirical) for TBL,
while $r_c=0.37$ (empirical) with $\gamma\equiv 1$ for channel.
Note that $\kappa=0.45$ (empirical) is universal in both
flows, as indicated by the linear slope.
Also included are $\kappa=0.40$ and $\kappa=0.38$, showing notable departure away from data.

Note that (\ref{eq:Ud2}) immediately yields a prediction of the mean velocity: $U^+=U/u_\tau=U^+_e-G/\kappa$.
Comparisons with data (with known empirical $U_e^+$) show impressive agreement, as seen in Fig.\ref{fig:Ell}.
The relative errors are bounded within $1\sim2\%$, at the same level as the data uncertainty.
The current description is as simple as the LPR (for channel) and much simpler than the MCN (for TBL involving
over ten free parameters), but with comparable if not better accuracy. Note also that the good description extends
up to $y^+_b= u_\tau y_b/\nu\approx50$ (the buffer layer thickness), which corresponds to
the breakdown of the quasi-balance condition near the wall (discussed elsewhere).

\begin{figure*}
\includegraphics[trim =4cm 1.5cm 3.8cm 2.0cm, clip, width = 5.5 cm]{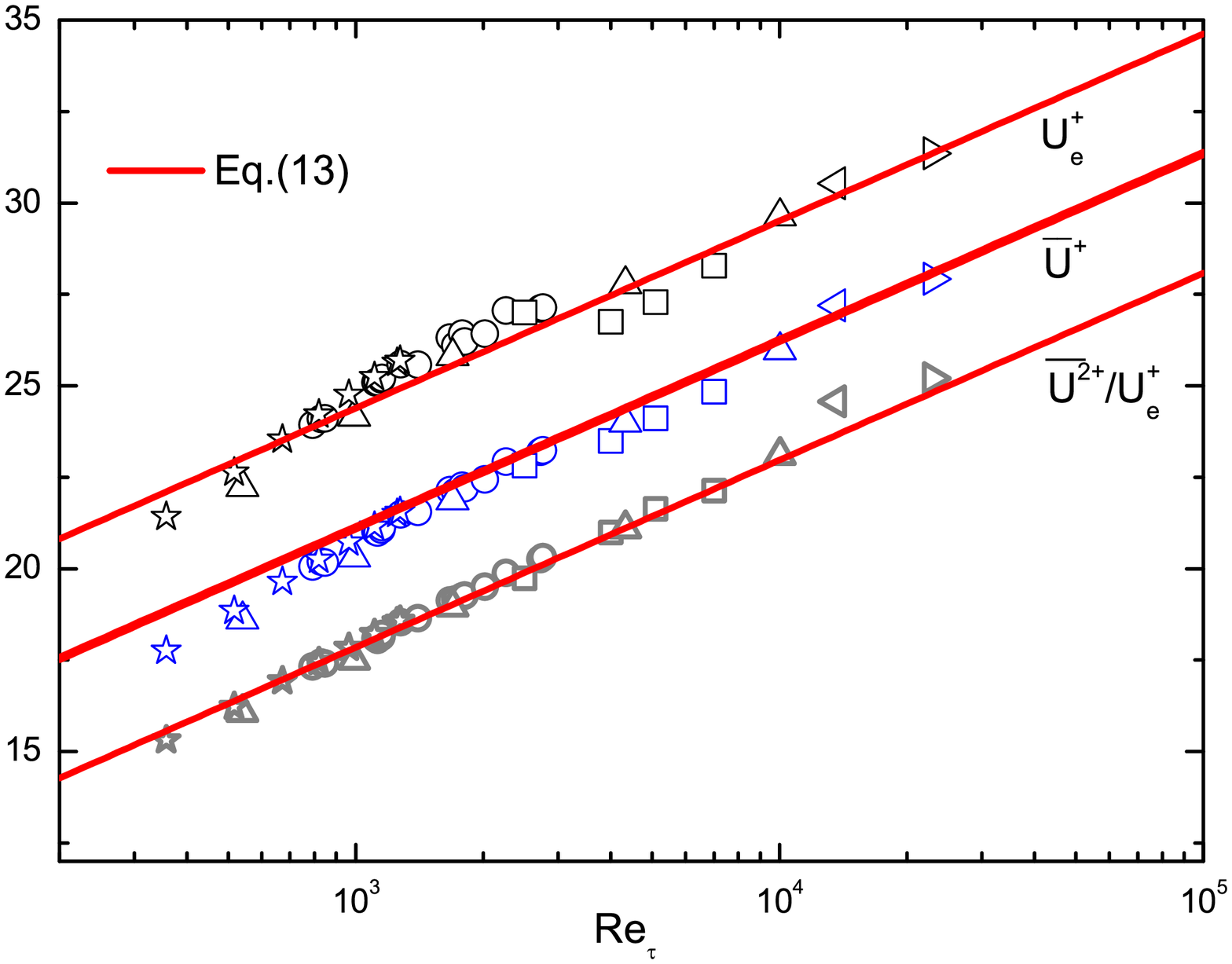}
\includegraphics[trim = 3.5cm 1.5cm 3.5cm 2.0cm, clip, width = 5.7 cm]{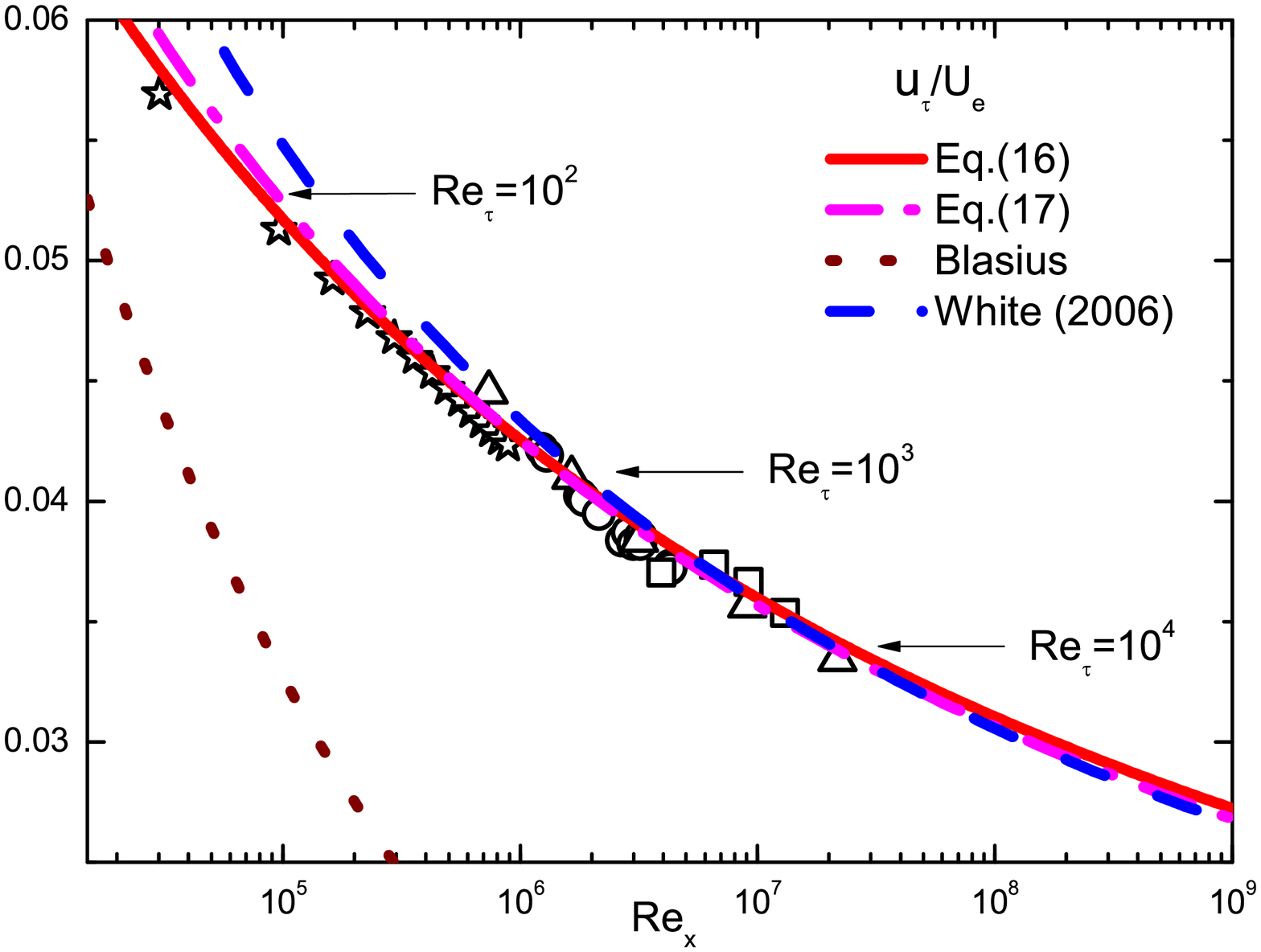}
\includegraphics[trim = 3.3cm 1.5cm 3.5cm 2.0cm, clip, width = 5.9 cm]{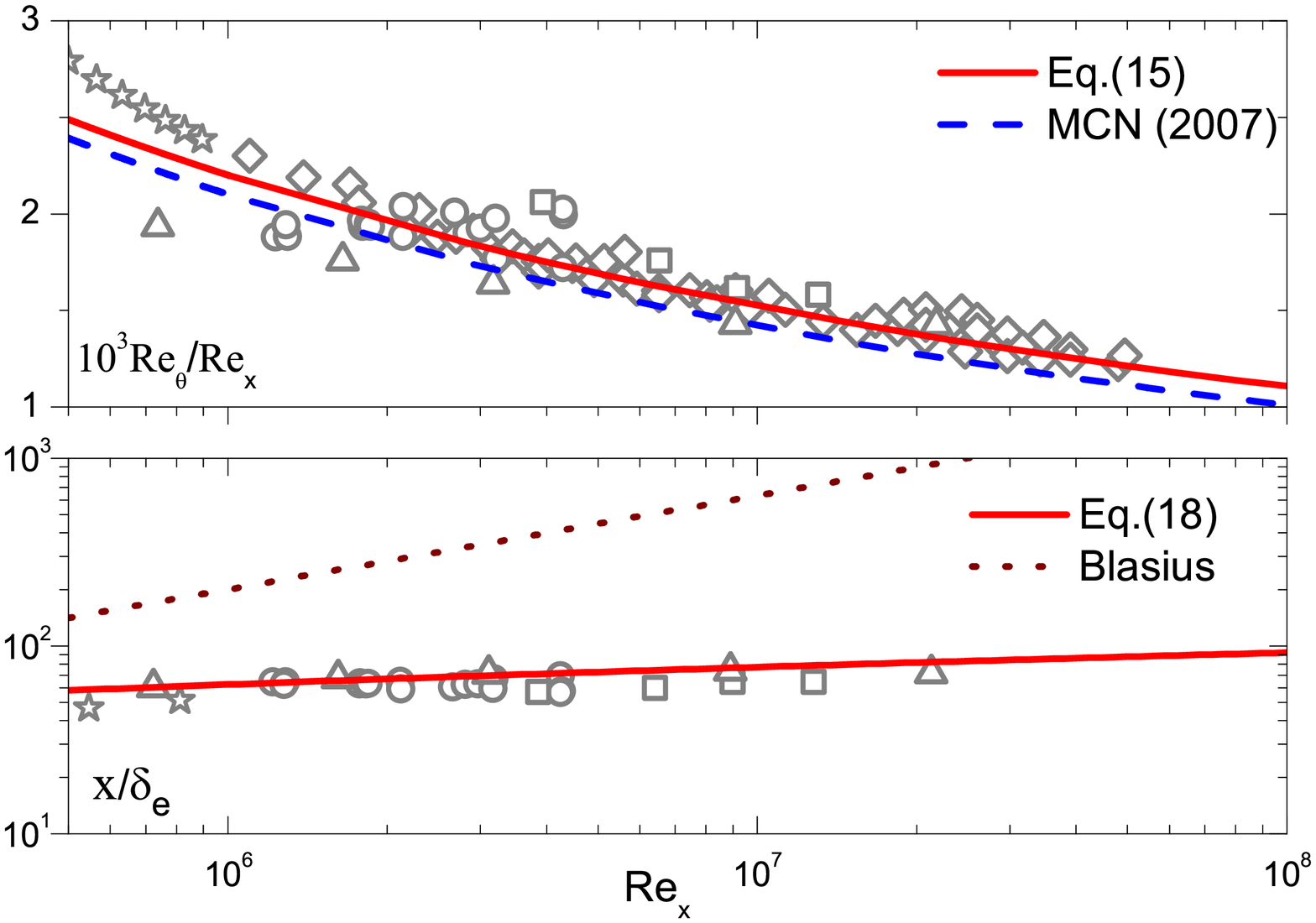}  
    \caption{(color online). [Left] Data of $U_e^+$ (black), $\overline{U}^+$ (blue) and $\overline{U^2}^+/ U_e^+$
    (gray) as functions of $Re_\tau$, compared with (\ref{eq:U}) (lines) showing good agreement. [Middle]
    Friction velocity as a function of ${{\mathop{\rm Re}\nolimits} _x}$ compared to the predictions ((\ref{utauxA}) and (\ref{eq:utau2}))
    (arrows depict the order of $Re_\tau$).
     [Right] The ratio of $10^3Re_\theta/Re_x$ (up panel) and $x/\delta_e$ (bottom panel), as a function of
     ${{\mathop{\rm Re}\nolimits} _x}$. In all panels, $\star$ are DNS data from \cite{schlatter2010}; all others are EXP data, i.e.
    $\bigtriangleup$ from \cite{Degraaff2000}, $\Box$ from \cite{Carlier}, $\triangleright$ from \citep{Nickels2007},
    $\bigcirc$ from \cite{Orlu2009}, $\triangleleft$ from \cite{Hutchins2009}, $\Diamond$ from Fig. 8 of \cite{monkewitz2007}.}
    \label{fig:Inter}
 \end{figure*}

The theory can predict streamwise development for relevant integral quantities in TBL. Three mean quantities are
estimated as a preparatory: $U_e$, $\overline{U}=\int_0^{1}{Udr}$ (vertically averaged mean velocity) and
$\overline{U^2}=\int_0^{1}{U^{2}dr}$ (second moment of $U$), which yields (see the Appendix):
\begin{eqnarray}
U_e^+  \approx \kappa ^{-1}\ln {{\mathop{\rm Re}\nolimits} _\tau }+ {B_e} \approx \ln {{\mathop{\rm Re}\nolimits} _\tau }/0.45 + 9.04 \nonumber\\
\overline{U}^+\approx  U^+_e-c_1 \approx \ln {\mathop{\rm Re}}_\tau/0.45+5.77  \nonumber\\
\overline{U^2}^+/ U_e^+\approx \overline{U}^+-c_1\approx \ln {\mathop{\rm Re}}_\tau/0.45+2.50 
\label{eq:U}\end{eqnarray}
Here, superscript $+$ indicates normalization with $u_\tau$, and $B_e\approx9.04$ is a constant involving wall
function calculation discussed elsewhere, while
$c_1=\kappa^{-1}\int_0^1G(r')dr'\approx3.27$
is theoretically determined
(see the Appendix). Comparison of (\ref{eq:U}) with data are illustrated in Fig.\ref{fig:Inter} (left) showing good agreement. Note that (\ref{eq:U}) implies that the familiar shape factor $H\equiv 1$, since the displacement thickness is defined by
${\delta^*} \equiv \int_0^{{\delta _e}} {(1 - U/{U_e})} dy$, thus $H\equiv{\delta^*}/\theta=
({{U_e^ +-\overline{U}^+}})/({{\overline{U}^+-\overline{U^2}^+/U^+_e}})=c_1/c_1\equiv 1$.
The higher order correction to (\ref{eq:U}) {and $H$} is discussed in the Appendix.

A new prediction is the ratio between the momentum thickness $Re$ ($Re_\theta=\theta U_e/\nu$) and the friction velocity $Re$
($Re_\tau=\delta_e u_\tau/\nu$) at the same $x$ (distance to the leading edge of a TBL). Substituting (\ref{eq:U}) into (\ref{alphafinal}), we obtain
\begin{eqnarray}\label{alphafinal}
\alpha \equiv {{\mathop{\rm Re}\nolimits} _{\theta} }/{{\mathop{\rm Re}\nolimits} _\tau }
=\overline{U}^+-\overline{U^2}^+/U^+_e \approx c_1\approx 3.27,
\end{eqnarray}
We can then connect ${{\mathop{\rm Re}\nolimits} _\theta }$ (and $Re_\tau$) with the streamwise Re
(${{\mathop{\rm Re}\nolimits} _x} = {U_e}x/\nu $), using (\ref{alphafinal}). Note that (\ref{eq:Cf}) indicates
${{d{{{\mathop{\rm Re}\nolimits} }_x}}}/{{d{{{\mathop{\rm Re}\nolimits} }_{\theta }}}}=dx/d\theta = {{U_e^{+2}}}$,
which, by noting
$U^+_e\approx\kappa^{-1}\ln (Re_\theta/\alpha)+B_e$ and integrating in $Re_\theta$ (or $U^+_e$) (dropping a higher order
term), leads to:
\begin{eqnarray}\label{RexRetheta}
  {{\mathop{\rm Re}\nolimits} _x}/{{\mathop{\rm Re}\nolimits} _{\theta} } &\approx &
  {(U_e^+ - 1/\kappa)^2 + \kappa^{-2}} \nonumber\\ &\approx &
  \kappa^{-2}[{(\ln ({\mathop{\rm Re}}_\theta/\alpha) + \kappa{B_e} - 1})^2 + 1] \nonumber\\ &\approx &
  4.94[{(\ln {{\mathop{\rm Re}\nolimits} _{\theta} } + 1.88)^2} + 1].
\end{eqnarray}
Compare to the MCN formula \cite{monkewitz2007}:
${{\mathop{\rm Re}\nolimits} _x}/{{\mathop{\rm Re}\nolimits} _\theta } = 6.7817{\ln ^2}{{\mathop{\rm Re}\nolimits} _\theta }
+ 3.6241\ln {{\mathop{\rm Re}\nolimits} _\theta } + 44.2971 + 50.5521/\ln {{\mathop{\rm Re}\nolimits} _\theta } + ...$,
(\ref{RexRetheta}) is much simpler (Fig. \ref{fig:Inter} right), and the coefficients are completely determined by three
parameters ($\kappa$, $\alpha$ and $B_e$).

Perhaps the most interesting prediction is the streamwise development of $u_\tau$. According to (\ref{eq:U}), $\ln {\mathop{\rm Re}\nolimits}_\tau=\kappa U_e/u_\tau- \kappa B_e$.
Since $Re_\theta=\alpha Re_\tau=\alpha \exp(\kappa U_e/u_\tau- \kappa B_e)$,
(\ref{RexRetheta}) thus further yields a useful relation:
\begin{eqnarray}\label{utauxA}
  {{\mathop{\rm Re}\nolimits} _x} \approx
  (\alpha/\kappa^2) \exp (\kappa {U_e}/{u_\tau } - \kappa{B_e})[{(\kappa {U_e}/{u_\tau } - 1)^2} + 1]
\end{eqnarray}
{Taking the logarithmic of (\ref{utauxA}), one has
\begin{eqnarray}\label{eq:utau2}
U_e/u_\tau&\approx&\kappa^{-1} \ln Re_x+B_e-\kappa^{-1}\ln(\alpha/\kappa^2)+O(\ln(\ln Re_x))\nonumber\\
&\approx& \ln Re_x/0.45+2.86-3.83\ln(\ln Re_x),
\end{eqnarray}
where the coefficient (-3.83) for $\ln(\ln Re_x)$ term is obtained using an exact solution of (\ref{utauxA})}:
$U_e/u_\tau\approx42$ at $Re_x=10^{10}$. Fig.\ref{fig:Inter} (right) shows (\ref{eq:utau2})
compared with data; the agreement is very satisfactory, notably better than the White's \cite{White2006} fitting
formula $U_e/u_\tau=\ln(0.06 Re_x)/0.477$ at moderate $Re_x$.

The streamwise growth of the boundary layer thickness is derived similarly.
From (\ref{alphafinal}), $\delta_e/\theta= U^+_e/\alpha\approx 0.68\ln Re_\tau+2.76$. Furthermore, using (\ref{RexRetheta}) and (\ref{eq:utau2}), we obtain
\begin{eqnarray}\label{eq:deltae}
x/\delta_e &=&\alpha Re_x /(U^+_e Re_\theta) \approx \alpha [U^+_e-2/\kappa+2/(\kappa^2 U_e^+)] \nonumber\\
&\approx& 7.27\ln Re_x-5.18-12.52\ln(\ln Re_x),
\end{eqnarray}
by neglecting the highest order term ($2/(\kappa^2 U_e^+)$).
This full analytic prediction is made for the first time, after Blasius's similar result
($x/\delta_{e} \approx 0.2 Re^{1/2}_x$) for a laminar boundary layer.
The agreement between data and (\ref{eq:deltae}) is generally good, within uncertainty in the definition
of the leading edge of a TBL due to the tripping force which introduces different virtual origins
(shifting the location of $x=0$) among different data sets in the calculation of
$Re_x$ \cite{monkewitz2007}, an issue to be explored in future.

Finally, the scaling of the vertical velocity at the boundary layer edge ($V_e$) can also be predicted.
Since $V_e=-\int_0^{\delta_e}\partial_x U dy=U_e (d\delta^\ast/dx)$ from (\ref{MME0}),
and $\delta^\ast=\delta_e(1-\overline{U}^+/U^+_e)=c_1\delta_e/U^+_e$ from (\ref{eq:U}),
one has (using (\ref{eq:utau2}) and (\ref{eq:deltae})):
\begin{eqnarray}\label{eq:Ve}
V_e/U_e =d\delta^\ast/dx \propto 1/\ln^2 {{\mathop{\rm Re}\nolimits} _x },
\end{eqnarray}
in contrast to $V_e/U_e\propto {Re^{-1/2}_x}$ in a laminar boundary layer.

In summary, a complete analytic theory for streamwise development of mean quantities ($u_\tau, \delta_e, \theta\ldots$) in TBL is
presented, agreeing well with data. The difference between internal channel and external TBL is
quantified, and the Karman constant is remarkably universal in the two flows.

We thank gratefully for many discussions with F. Hussain. This work is supported by National Natural
Science (China) Fund 11452002 and 11521091.

\appendix
\emph{Appendix.}
Note that $U_e^+-\overline{U}^+=\int_0^1U^+_ddr'$. Assuming (\ref{eq:Ud2}) extends to the wall, the near-wall contribution
up to the buffer layer is estimated to be $O(\ln Re_\tau/Re_\tau)$ and is neglected; we thus have
$\int_0^1U^+_ddr'\approx c_1$ where $c_1=\kappa^{-1}\int_0^{1}G(r')dr'$,
with $G(r')=\int_0^{r'}f(\hat{r})d\hat{r}$ in (\ref{eq:Ud2}). We further rewrite
$G(r')=\int_0^{r'}(1-\hat{r})^{-1}d\hat{r}+ \int_0^{r'}g(\hat{r})d\hat{r}$, where $g(\hat{r})=f(\hat{r})-(1-\hat{r})^{-1}$
is a smooth function bounded in the domain $0\leq\hat{r}\leq1$, to remove a singularity
$f(\hat{r})\propto (1-\hat{r})^{-1}$ at $\hat{r}=1$ and to obtain a finite $g$. Since
$\int_0^1\int_0^{r'}(1-\hat{r})^{-1}d\hat{r}dr'=1$, $\int_{0}^1 \int_0^{r'} {g(\hat{r})d\hat{r}} dr'\approx0.47$
(numerical integration), we thus have $c_1=\kappa^{-1}(1+0.47)\approx3.27$. Similarly,
$\int_0^1U^{+2}_ddr=U_e^{+2}+\overline{U^2}^+-2U_e^+\overline{U}^+\approx c_2$ where $c_2=\kappa^{-2}\int_0^1G^2(r')dr'\approx19.55$.
It leads to $\overline{U^2}^+/U^+_e\approx2\overline{U}^+-U_e^{+}+c_2/U^+_e\approx\overline{U}^+-c_1+c_2/U^+_e$,
and $H=c_1/(c_1-c_2/U^+_e)> 1$ for finite $Re$ ($H\approx1.26$ at $Re_\tau=10^4$). Finally, let us calculate $B_e$.
As (\ref{eq:Ud2}) extends to $y^+_b$ (buffer layer thickness), then
$U^+_e=U^+_b+\kappa^{-1}\int_0^{1-y^+_b/Re_\tau}f(\hat{r})d\hat{r}$,
which yields $U_e^+=\ln Re_\tau /\kappa+ B_e$, where
$B_e=(U^+_b- \kappa^{-1}\ln y^+_{b})+\kappa^{-1}\int_0^{1-y^+_b/Re_\tau}g(\hat{r})d\hat{r}$, determined by
$y^+_b$ and $U^+_b$ (its precise values will be discussed elsewhere). Here, we let $B_e=9.04$ to fit all data in
figure \ref{fig:Inter} (left), which is the only fitting parameter to yield (\ref{eq:U}).


\begin{thebibliography}{99}

\bibitem{Voyage}   P. A. Davidson, et al., {\it A voyage through turbulence}. (Cambridge University Press, 2011).

\bibitem{SmitsMarusic2013} A.J. Smits \& I. Marusci, {Physics Today}, 66(9), 25 (2013).

\bibitem{Prandtl} L. Prandtl, L., Third International Mathematical Congress, Heidelberg.
484-491 (1904). L. Prandtl, Z. Angew. Math. Mech. 5, 136-139 (1925).

\bibitem{Reynolds} Reynolds, O., Phil. Trans. R. Soc. London, Ser. A. 174, 935-982 (1883).

\bibitem{Marusic2010} I. Marusic, \emph{et al.}, {Phys. Fluids.} {22}, 065103 (2010).

\bibitem{Yakhot2010} V. Yakhot, {Phys. Rev. E} {82}, 045301(R) (2010).

\bibitem{millikan1938} C. B. Millikan, In {\it Proceedings 5th International Congress on Applied Mechanics}. Cambridge, MA 386-392 (1938).

\bibitem{Barenblatt} G.I. Barenblatt \& A.J. Chorin, {Proc. Natl Acad. Sci.} USA 101, 15023-15026 (2004).

\bibitem{George06} W. K. George, AIAA Journal, 44(11), 2435-2449 (2006).

\bibitem{Nickels04} T. B. Nickels, J. Fluid Mech. 521, 217-239 (2004).

\bibitem{monkewitz2007} P. A. Monkewitz, et al., {Phys. Fluids.} 19, 115101 (2007).

\bibitem{Lvov2008} V.S. L'vov, \emph{et al.}, {Phys. Rev. Lett.} {100}, 050504 (2008).

\bibitem{shenjp} Z.S. She, \emph{et al.}, New Journal of Physics, 14 093054 (2012).

\bibitem{Jimenez2012} J. Jimenez, {Annu. Rev. Fluid Mech.} {44}, 27-45 (2012).

\bibitem{Degraaff2000} D. B. Degraaff \& J. K. Eaton, Journal of Fluid Mechanics, 422, 319�346 (2000).

\bibitem{Jimenez2010} J. Jimenez, \emph{et al.}, Journal of Fluid Mechanics 657, 335�360 (2010).

\bibitem{schlatter2010} P. Schlatter, et al., International Journal of Heat and Fluid Flow, 31(3), 251-261 (2010).

\bibitem{Monty2005} J. P. Monty, Developments in smooth wall turbulent duct flows. PhD Thesis, University of Melbourne (2005).

\bibitem{Carlier} J. Carlier, \& M. Stanislas, Journal of Fluid Mechanics, 535, 143�188 (2005).

\bibitem{Nickels2007} T. B. Nickels, et al., Philosophical transactions of The Royal Society A 365, 807�822. (2007).

\bibitem{Orlu2009} R. Orlu, Experimental studies in jet flows and zero pressure-gradient turbulent boundary
layers. PHD thesis, KTH, Stockholm (2009).

\bibitem{Hutchins2009} N. Hutchins, et al., Journal of Fluid Mechanics, 635, 103�136 (2009).




\bibitem{White2006} F. M. White, 2006 {\it Viscous fluid flow}, 3rd edn. New York, NY: McGraw-Hill.








\end{thebibliography}
\end{document}